\documentclass[preprint,12pt]{elsarticle}

\usepackage{amssymb}
\usepackage{amsmath}

\usepackage{booktabs} 
\usepackage{makecell}
\usepackage{float}

\usepackage[ruled,vlined]{algorithm2e}
\usepackage[section]{placeins}

\journal{Applied Thermal Engineering}

\begin{document}

\begin{frontmatter}



\title{Generalised Mixing-Plane Method for Compressible Reacting-Mixture Flows in Steady Multiphysics Turbomachinery Simulations}


\author[label1]{Yifeng Wang},
\author[label1]{Lin Shi},
\author[label2]{Fenglai Huang},
\author[label1]{Rui Wang}
\author[label1]{Feng Wang\corref{cor1}}
\ead{feng.wang@sjtu.edu.cn}
\author[label1]{Hui Xu\corref{cor1}}
\ead{dr.hxu@sjtu.edu.cn}

\cortext[cor1]{Corresponding author.}

\affiliation[label1]
            {organization={School of Aeronautics and Astronautics, Shanghai Jiao Tong University},
            city={Shanghai},
            postcode={200240}, 
            country={China}}
\affiliation[label2]
            {organization={Wuhan Second Ship Design and Research Institute},
            city={Wuhan},
            postcode={430205}, 
            country={China}}

\begin{abstract}
The mixing-plane method plays a pivotal role in steady simulations of multiple turbomachinery components. With advances in computational power, a whole-engine gas turbine  simulation is no longer an elusive approach. However, whole-engine simulations require simultaneous coupling of the compressor, turbine and combustor. A key challenge is that the working fluid after the combustor can no longer be treated as a perfect gas, since the combustion introduces composition variations and combustion products. Conventional mixing-plane methods based on a fixed-composition gas cannot guarantee a thermodynamically consistent mixed-out state. To overcome this difficulty, this paper extends the classic mixing-plane approach to handle compressible reacting flows. The nonlinearity in the thermodynamic closure is addressed using a novel nested algorithm combining an outer pressure-root search with an inner enthalpy inversion. Physical state checks and power-law pressure sampling are incorporated to improve numerical robustness at low normal velocities where the pressure root approaches its upper physical bound. The performance of the method is demonstrated in three configurations: a quasi-1D interface test elucidating the thermodynamic stiffness of the proposed mixing-plane formulation for different fuel types and flow conditions; the Darmstadt transonic compressor case that verifies the proposed method reduces to the classic mixing-plane formulation when the mixture fraction reduces to zero; and a whole-engine simulation of the KJ66 micro-turbojet is used to demonstrate the performance of the method for multiphysics turbomachinery simulations. Good agreement with experimental data is observed, and the maximum relative mass-flow error across the mixing planes remains below $0.11\%$, demonstrating conservative and thermodynamically consistent interface coupling for compressible reacting turbomachinery flows.

\end{abstract}


\begin{highlights}
\item A generalised mixing-plane method for compressible reacting flow is developed.
\item Robust pressure-root recovery at near-zero normal velocity is demonstrated.
\item A thermochemical indicator for enthalpy-inversion stiffness is characterised.
\item The non-reacting limit is verified against the classical mixing-plane method.
\item Whole-engine simulations of KJ66 MTE show good agreement with experimental data.
\end{highlights}

\begin{keyword}
Mixing-plane method \sep whole-engine simulation \sep compressible reacting flow \sep steady turbomachinery simulation


\end{keyword}

\end{frontmatter}


\section{Introduction}

Gas turbine engines are highly integrated systems in which the aero-thermal performances of individual components are strongly coupled. Multistage simulations can be insufficient to capture the interactions among engine components. Extensive hardware builds and tests are still required to test the performance of engines under different operating conditions. High-fidelity whole-engine computational fluid dynamics (CFD) simulations provide valuable insight into component interactions, thereby supporting more informed design and performance evaluation while potentially reducing engine design costs and shortening development cycles \cite{sandberg2022fluiddynamics,briones2021fully, wang2016virtual, xu2025full}. Unsteady simulations can explicitly resolve the temporal interactions between adjacent blade rows \cite{erdos1977periodicfan,giles1990statorrotor,pichler2018lesgap,zhu2022high}, but their computational costs are generally prohibitive at the design stage. Although several computationally efficient unsteady formulations have been developed \cite{he1998efficientapproach,hall2002harmonicbalance,wang2022fourier}, their application to whole-engine simulations, especially related to the combustion process, remain to be dubious \cite{tucker2011computation}. Steady simulations based on the mixing-plane method remain the workhorse for turbomachinery simulations, as they can account for component interactions with manageable computational effort while retaining satisfactory accuracy \cite{du2016novelmixingplane,cornelius2013steadytransientcompressor}. In such simulations, adjacent blade rows or engine components are often solved in separate computational domains. The accuracy of the coupled calculation then depends not only on the solution within each domain, but also on the conservative and physically consistent transfer of mass, momentum, energy and transported scalars across their interfaces.

The mixing-plane method, which serves as the standard interface treatment for multistage turbomachinery computation, was established by Denton \cite{denton1992multistage}. It replaces the circumferentially non-uniform flow exchanged between adjacent blade rows with an equivalent mixed-out state, thereby removing periodic blade-row unsteadiness while retaining the mean inter-row coupling. Conservative formulations based on flux averaging further established that the mixed-out state should reproduce the integral mass, momentum, and energy fluxes of the incoming non-uniform profile, rather than simply average primitive variables \cite{cumpsty2006averaging,frey2023averagingtechniques}. Subsequent developments have improved the robustness \cite{wang2014improvedmixingplane}, implicit coupling \cite{hanimann2014fullyimplicitmixingplane,xu2022conservativemixingplane}, reverse-flow treatment \cite{du2016novelmixingplane,wang2014improvedmixingplane}, non-reflecting behaviour \cite{du2016novelmixingplane,gisbert2016nonreflectingmixingplane}, and adjoint/design capability of mixing-plane interfaces \cite{rodrigues2018adjoint,vitale2020su2multistage}. Nevertheless, these formulations are predominantly constructed for fixed-composition gases, in which the thermodynamic state can be recovered using a perfect-gas closure.

At combustor--turbine interfaces, the transferred flow may contain combustion products with non-uniform composition. For fixed-composition gases, conventional mixing-plane methods recover the mixed-out state from conservative fluxes using a prescribed perfect-gas closure \cite{ray2015robustmixingplane}; similar formulations have also been extended to fixed-composition non-ideal fluids in turbomachinery \cite{vitale2020su2multistage}. Variable-composition mixtures introduce an additional difficulty: species or mixture scalars must be conserved, and thermophysical properties must be evaluated from the reconstructed composition \cite{northall2006variablegas,saghafian2015compressibleflamelet}. Otherwise, the fluxes may be conserved while the recovered state remains thermodynamically inconsistent. The challenge is therefore to ensure both flux conservation and composition-dependent thermodynamic consistency.


It should be noted that the mixing-plane formulation for incompressible flows can be greatly simplified, because energy is decoupled from mass and momentum, allowing temperature and pressure to be evaluated independently. In compressible reacting flow, however, flow velocities, temperature, pressure and composition are closely coupled with one another; this introduces significant nonlinearity in the numerical procedure to work out a mixed-out state. To the authors’ best knowledge, no conservative mixing-plane formulation has been reported in the open literature capable of handling compressible reacting flows.

The objective of this paper is to develop a flux-conservative mixing-plane formulation for steady compressible reacting-mixture simulations. The proposed method extends the classic mixing-plane formulation from fixed-composition gas flows to variable-composition reacting-mixture flows. It conserves the interface fluxes of mass, three-component momentum, total enthalpy, and scalar transport, and recovers pressure, temperature, density, velocity, and thermophysical properties from a single thermodynamically consistent reacting-mixture state. A pressure-based recovery procedure with admissibility checks is introduced to obtain a physically admissible mixed-out state under the reacting-mixture closure. The formulation is evaluated using a quasi-one-dimensional theoretical interface test that isolates the nonlinear recovery process, a Darmstadt transonic compressor case that verifies the behaviour of reduction to the conventional non-reacting conservative mixing plane, and a KJ66 micro turbine engine simulation that demonstrates the method in a coupled reacting whole-engine calculation.

This paper is organised as follows: Section~\ref{sec:method} explains the detailed mathematical derivation of the proposed mixing-plane method. Section~\ref{sec:computational_framework} introduces the computation framework used in this study. Section~\ref{sec:results} presents the theoretical, compressor, and whole-engine test cases. Section~\ref{sec:conclusions} draws the main conclusions.

\section{Methodology}
\label{sec:method}

\subsection{Mixing-plane method revisited}
\label{subsec:classical_formulation}

\begin{figure}[htbp]
\centering
\includegraphics[width=\columnwidth]{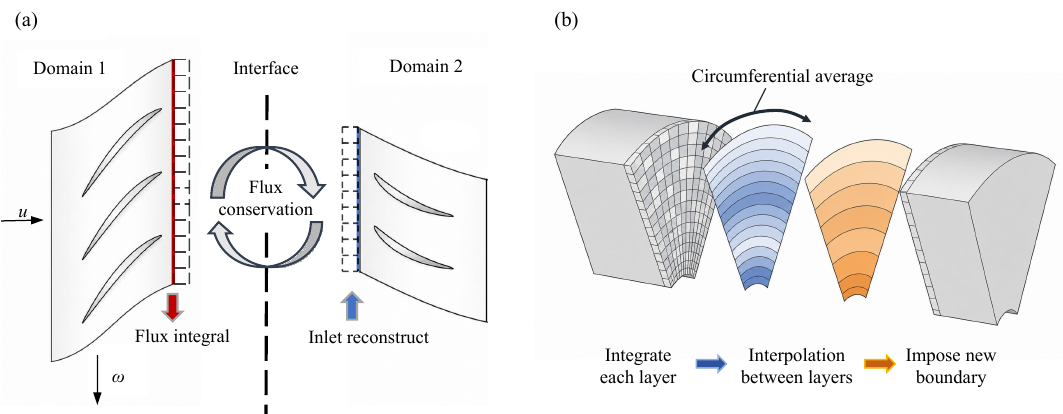}
\caption{Schematic of the conservative mixing-plane treatment: (a) flux integration and mixed-state reconstruction across adjacent domains; (b) circumferential averaging and radial interpolation between blade-row interfaces, reproduced from \cite{hanimann2014fullyimplicitmixingplane}.}
\label{fig:schematic_view}
\end{figure}

In steady turbomachinery simulations, adjacent blade rows or engine components are commonly solved in separate computational domains and coupled through a mixing-plane interface. As illustrated in Figure~\ref{fig:schematic_view}(a), the flow variables at the upstream side of the interface are first circumferentially integrated to obtain the corresponding conserved fluxes. Then a mixed-out state is reconstructed from these fluxes and imposed as the inlet condition for the downstream domain. As shown in Figure~\ref{fig:schematic_view}(b), grid cells within the same radial band are grouped into an annular layer over which the conserved fluxes are circumferentially integrated. The fidelity of the mixing-plane treatment therefore depends on whether the reconstructed state preserves these integral fluxes while satisfying the thermodynamic closure at the interface.

\FloatBarrier


In classical formulations, without loss of generality, let $x$ denote the direction normal to the mixing plane and $y$ the pitchwise direction. The basic principle of the mixing-plane approach can then be illustrated using the two-dimensional Euler equations:

\begin{equation}
\frac{\partial \mathbf{U}}{\partial t}
+
\frac{\partial \mathbf{F}}{\partial x}
+
\frac{\partial \mathbf{G}}{\partial y}
=0, 
\end{equation}
where the vector of conservative variables and the inviscid flux vectors
are given by

\begin{equation}
\mathbf{U}
=
\begin{bmatrix}
\rho \\
\rho u \\
\rho v \\
\rho E
\end{bmatrix},
\qquad
\mathbf{F}
=
\begin{bmatrix}
\rho u \\
\rho u^2+p \\
\rho uv \\
\rho u h_t
\end{bmatrix},
\qquad
\mathbf{G}
=
\begin{bmatrix}
\rho v \\
\rho uv \\
\rho v^2+p \\
\rho v h_t
\end{bmatrix}.
\label{eq:euler_vectors}
\end{equation}
Here, $\rho$ is the density, $u$ and $v$ are the velocity components in
the $x$- and $y$-directions, respectively, $p$ is the static pressure,
$E$ is the total specific energy, and $h_t$ is the total specific
enthalpy.

At the mixing plane, the fluxes normal to the interface are averaged over
one blade pitch $\Delta y$. Because of pitchwise periodicity, the net
contribution of $\mathbf{G}$ vanishes upon integration over one blade
pitch. The pitch-averaged normal fluxes are therefore defined as

\begin{equation}
\label{eq:integrate_flux}
\begin{aligned}
f_m
&=
\frac{1}{\Delta y}
\int_{y_0}^{y_0+\Delta y}
\rho u\mathrm{d}y,\\
f_{Mx}
&=
\frac{1}{\Delta y}
\int_{y_0}^{y_0+\Delta y}
\left(\rho u^2+p\right)\mathrm{d}y,\\
f_{My}
&=
\frac{1}{\Delta y}
\int_{y_0}^{y_0+\Delta y}
\rho uv\mathrm{d}y,\\
f_h
&=
\frac{1}{\Delta y}
\int_{y_0}^{y_0+\Delta y}
\rho u h_t\mathrm{d}y.
\end{aligned}
\end{equation}
Here, $y_0$ denotes the lower bound of the pitchwise integration interval,
while $f_m$, $f_{Mx}$, $f_{My}$, and $f_h$ denote the pitch-averaged mass,
$x$-momentum, $y$-momentum, and total-enthalpy fluxes, respectively.

Together with the specific total enthalpy
\begin{equation}
\label{eq:total_enthalpy_method1}
    h_t=\frac{\gamma}{\gamma-1} \frac{p}{\rho}+\frac{1}{2} (u^2 + v^2),
\end{equation}
where $\gamma$ denotes the ratio of specific heats and is assumed constant for a fixed-composition perfect gas, the system of Eq.~(\ref{eq:integrate_flux}) can be solved \cite{giles1991unsflo} to obtain
\begin{equation}
\label{eq:giles_sol}
\begin{aligned}
    \bar{\rho} &=\frac{f_m}{\bar{u}},\\
    \bar{u} &= \frac{f_{Mx} - \bar{p}}{f_m},\\
    \bar{v} &= \frac{f_{My}}{f_m},\\
    \bar{p} &= \frac{1}{\gamma+1} \left[f_{Mx} + \sqrt{\gamma^2f_{Mx}^2 + (\gamma^2-1)(f_{My}^2-2f_mf_h)} \right].\\
\end{aligned}
\end{equation}

This mixed-out state is subsequently imposed as the average inflow condition for the downstream domain. 
The above formulation is straightforward for a fixed-composition perfect gas, for which the mixed-out state can be recovered directly from the conserved fluxes.

For reacting mixtures, however, a unique challenge emerges: enthalpy depends on both temperature and mixture composition. Therefore, pressure can no longer be solved via a quadratic equation from Eq.~(\ref{eq:giles_sol}), as $\gamma$ is no longer constant. Instead, the additional scalar fluxes providing composition information must be integrated into the state recovery procedure to account for variable-property thermodynamics.

\subsection{Generalised formulation for compressible reacting flow}
\label{subsec:governing_equations}
The preceding subsection shows that, for a fixed-composition perfect gas, the mixed-out state can be recovered explicitly from the averaged mass, momentum, and total enthalpy fluxes. For a reacting mixture, additional information is required to determine the local composition and the corresponding thermodynamic properties. In the present implementation, the mixture fraction $Z\in[0,1]$ is used as the sole composition coordinate, such that the species mass fractions are represented by $\boldsymbol{Y}=\mathcal{Y}(Z)$ \cite{peters2001turbulent}. This relation determines the mixture composition but does not prescribe the temperature, for compressible reacting flow, which must be recovered consistently from the conserved enthalpy flux \cite{saghafian2015compressibleflamelet}.

\begin{figure}[t]
\centering
\includegraphics[width=0.55\columnwidth]{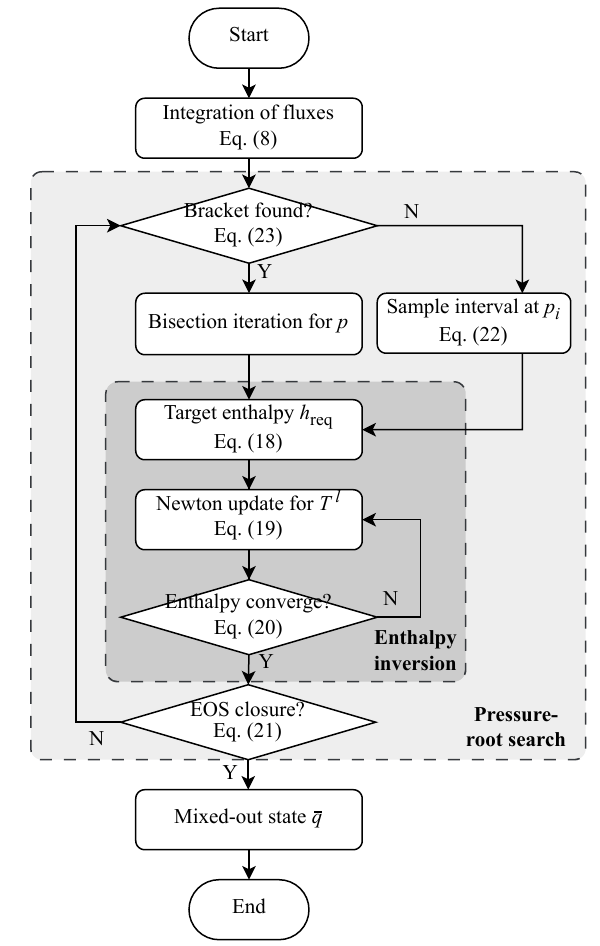}
\caption{Overall framework of the flux-based mixed-out state reconstruction with an outer pressure search and an inner enthalpy-to-temperature inversion.}
\label{fig:flow_chart}
\end{figure}

Let $A_j$ denote the area of the $j$th annular averaging band, with the band index omitted below for clarity. Let $\boldsymbol{n}$ be its unit normal vector. For a local state
$\boldsymbol{q}=(\rho,\boldsymbol{u},p,T,Z)$, the normal velocity and specific total enthalpy are defined as
\begin{equation}
    u_n=\boldsymbol{u}\cdot\boldsymbol{n},
    \qquad
    h_t=h(T,Z)+\frac{1}{2}\boldsymbol{u}\cdot\boldsymbol{u},
\label{eq:un and ht}
\end{equation}
where $h$ is static enthalpy. The inviscid flux normal to the interface is then
\begin{equation}
    \boldsymbol{\Phi}_n(\boldsymbol{q})=
    \begin{bmatrix}
    \rho u_n \\
    \rho u_n\boldsymbol{u}+p\boldsymbol{n} \\
    \rho u_n h_t \\
    \rho u_n Z
    \end{bmatrix}.
\end{equation}

The corresponding area-averaged flux vector supplied to the mixed-out reconstruction is
\begin{equation}
\boldsymbol{\mathcal{F}}
=
\frac{1}{A}
\int
\boldsymbol{\Phi}_n(\boldsymbol{q})
\mathrm{d}A
\equiv
\begin{bmatrix}
f_m \\
\boldsymbol{f}_M \\
f_{ht} \\
f_Z
\end{bmatrix},
\label{eq}
\end{equation}
where $f_m$, $\boldsymbol{f}_M$, $f_{ht}$, and $f_Z$ denote the averaged mass, three-component momentum, total enthalpy, and mixture fraction fluxes, respectively. Given these averaged fluxes, the objective is to recover a uniform mixed-out state
$\overline{\boldsymbol{q}}
=(\overline{\rho},\overline{\boldsymbol{u}},
\overline{p},\overline{T},\overline{Z})$
whose normal flux reproduces the conservation values:
\begin{equation}
    \boldsymbol{\Phi}_n\left(\overline{\boldsymbol{q}}\right)
    =
    \boldsymbol{\mathcal{F}}.
\label{eq:target}
\end{equation}

Thus, $\boldsymbol{\mathcal{F}}$ is the known input to the reconstruction, whereas $\overline{\boldsymbol{q}}$ is the mixed-out state to be recovered. As illustrated in Figure~\ref{fig:flow_chart}, the reconstruction employs an outer pressure search with an embedded enthalpy-to-temperature inversion to obtain a state that satisfies both the prescribed fluxes and the thermodynamic closure. The required thermochemical relations are introduced in Section~\ref{subsec:mixed_state_thermo}, followed by the detailed recovery procedure in Section~\ref{subsec:pressure_recovery}.

\FloatBarrier

\subsection{Thermochemical closure of the reacting-mixture state}
\label{subsec:mixed_state_thermo}
As indicated by Eq.~(\ref{eq:total_enthalpy_method1}), the classical mixed-out reconstruction relies on prescribed gas properties to relate total enthalpy to pressure, density, and velocity. In a compressible reacting mixture, however, the specific gas constant $R$ and heat capacity ratio $\gamma$ vary with mixture composition and temperature and therefore cannot be treated as prescribed constants. State recovery additionally requires a consistent relation among mixture fraction, species composition, temperature, and thermodynamic properties. A thermochemical closure that maps $Z$ and $T$ to the species composition, mixture gas constant, and the corresponding caloric properties, must therefore be established before the mixed-out state can be recovered.

For a prescribed mixture fraction $Z$, the species mass fractions are obtained from the same one-dimensional composition manifold employed in the neighbouring computational domains:
\begin{equation}
Y_k
=
\mathcal{Y}_k(Z),
\qquad
k=1,\ldots,N_s,
\qquad
\sum_{k=1}^{N_s}Y_k=1.
\label{eq:species_composition}
\end{equation}
Here $N_s$ is the number of species. The mixture molecular weight and specific gas constant are then evaluated as
\begin{equation}
\frac{1}{W(Z)}
=
\sum_{k=1}^{N_s}
\frac{Y_k(Z)}{W_k},
\qquad
R(Z)
=
\frac{R_u}{W(Z)},
\label{eq:mixture_gas_constant}
\end{equation}
where $W_k$ is the molecular weight of species $k$ and $R_u$ is the universal gas constant.

The molar-specific species enthalpy $h_k(T)$ and heat capacity $c_{p,k}(T)$ are evaluated from the NASA 7-coefficient polynomial \cite{mcbride2002nasa}, 
\begin{equation}
\label{eq:janaf_enthalpy_method}
     \frac{h_k^{\mathrm{mol}}(T)}{R_uT}
     =a_{k,1}+a_{k,2}\frac{T}{2}+a_{k,3}\frac{T^2}{3}
     +a_{k,4}\frac{T^3}{4}+a_{k,5}\frac{T^4}{5}
     +\frac{a_{k,6}}{T},
 \end{equation}
 \begin{equation}
\label{eq:janaf_cp_method}
    \frac{c_{p,k}^{\mathrm{mol}}(T)}{R_u}
    =a_{k,1}+a_{k,2}T+a_{k,3}T^2+a_{k,4}T^3+a_{k,5}T^4.
\end{equation}
The mass-specific mixture enthalpy and heat capacity are:
\begin{equation}
\label{eq:mixture_enthalpy_cp_method}
    h(T,Z)=\sum_{k=1}^{N_s}Y_k(Z)\frac{h_k^{\mathrm{mol}}(T)}{W_k},
    \qquad
    c_p(T,Z)=\sum_{k=1}^{N_s}Y_k(Z)\frac{c_{p,k}^{\mathrm{mol}}(T)}{W_k}.
\end{equation}

The thermal equation of state is
\begin{equation}
p
=
\rho R(Z)T.
\label{eq:mixture_equation_of_state}
\end{equation}

Together, these thermochemical relations enforce consistency among pressure, density, temperature, and mixture composition, thereby providing the basis for recovering a physically admissible mixed-out state from the area-averaged conservative fluxes. On this basis, the following subsection develops a pressure-based procedure for recovering the mixed-out state.

\subsection{Pressure-based recovery of the mixed-out state}
\label{subsec:pressure_recovery}

With the thermochemical closure established, the mixed-out variables do not need to be solved simultaneously. Instead, the conservation relations can be used to express the reconstructed state in terms of a single trial pressure. For a nonzero averaged mass flux, $f_m\neq0$, the mixture fraction and specific total enthalpy follow directly from the scalar and total enthalpy fluxes:
\begin{equation}
    \overline{Z}=\frac{f_Z}{f_m},
    \qquad
    \overline{h_t}=\frac{f_{ht}}{f_m}.
\label{eq:barZ barH}
\end{equation}
Both quantities are fixed by the prescribed fluxes and remain unchanged during the subsequent pressure solution.

For a trial pressure $p$, the momentum flux constraint determines the velocity vector. The normal velocity and density then follow from the velocity definition and mass flux constraint:
\begin{equation}
    \boldsymbol{u}(p)=\frac{\boldsymbol{f}_M-p\boldsymbol{n}}{f_m},
    \qquad
    u_n(p)=\boldsymbol{u}(p)\cdot\boldsymbol{n},
    \qquad
    \rho(p)=\frac{f_m}{u_n(p)}.
\label{eq:u un rho}
\end{equation}
Here, $\boldsymbol{u}(p)$ and $\rho(p)$ denote temporary states associated with the trial pressure; overbars are reserved for the final mixed-out state.

The static enthalpy required by the total enthalpy constraint is
\begin{equation}
    h_{\mathrm{req}}(p)=\overline{h_t}-\frac{1}{2}\boldsymbol{u}(p)\cdot\boldsymbol{u}(p).
\label{eq:h_req}
\end{equation}
For each trial pressure, the corresponding temperature is obtained by matching this required enthalpy. The corresponding temperature $T(p)$ is then recovered from the thermochemical closure by solving $h(T,\overline{Z})=h_{\mathrm{req}}(p)$. The temperature residual and corresponding Newton update are defined as:
\begin{equation}
    T^{(\ell+1)} = T^{(\ell)}-\frac{h\left(T^{(\ell)},\overline{Z}\right)-h_{\mathrm{req}}(p)}{c_p\left(T^{(\ell)},\overline{Z}\right)},
\label{eq:newton_t}
\end{equation}
\begin{equation}
    \mathcal{R}_T = h(T,\overline{Z})-h_{\mathrm{req}}(p) \rightarrow 0,
\label{eq:res_t}
\end{equation}
where $\ell$ is the temperature iteration index. Once the enthalpy residual converges, the resulting temperature is denoted by $T(p)$. Because $h(T,\overline{Z})$ is monotonic within the valid thermochemical range, this inversion yields a unique temperature whenever $h_{\mathrm{req}}(p)$ is thermodynamically admissible.

At this stage, the conservation constraints have determined $\boldsymbol{u}(p)$, $\rho(p)$, $h_{\mathrm{req}}(p)$, and $T(p)$ for the prescribed pressure. The remaining requirement is consistency with the equation of state. This condition is expressed by the scalar pressure residual
\begin{equation}
    \mathcal{R}_p(p)=p-\rho(p)R(\overline{Z})T(p),
    \qquad
    \mathcal{R}_p(\overline{p})\rightarrow0,
\label{eq:res_p}
\end{equation}
where $\overline{p}$ is the recovered mixed-out pressure. Once this root is obtained, the remaining primitive variables are evaluated as
$\overline{\boldsymbol{u}}=\boldsymbol{u}(\overline{p})$,
$\overline{\rho}=\rho(\overline{p})$, and
$\overline{T}=T(\overline{p})$.

The coupled mixed-state reconstruction has therefore been reduced to a scalar pressure equation with an embedded enthalpy-to-temperature inversion. For every valid trial pressure, the mass, momentum, total enthalpy, and mixture fraction flux constraints are satisfied by construction; the pressure residual measures the remaining inconsistency with the equation of state.

It should be noted that the pressure recovery becomes ill-conditioned when the mixed normal velocity approaches zero. From Eq.~(\ref{eq:u un rho}), a vanishing normal velocity ${u}_n(p) \to 0$ implies that the pressure root approaches the upper admissible bound $p_{\max}=\mathbf f_M\cdot\mathbf n$. In this limit, the density expression becomes singular and the pressure residual may change sign only within a very narrow interval adjacent to $p_{\max}$. The treatment of roots close to $p_{\max}$ is therefore included in the admissible pressure-search procedure described below.

\subsection{Admissible pressure root search at $u_n \approx 0$}
\label{subsec:admissible_pressure_search}

Following the limiting behaviour identified in the preceding subsection, the singular point $p_{\max}$ is excluded from the pressure search. The numerical interval is defined by $p_{\mathrm{lo}}=\delta_p$ and $p_{\mathrm{hi}}=p_{\max}-\delta_p$, where the small positive pressure offset $\delta_p$ excludes both the non-positive lower limit and the singular upper bound. Before the pressure search, the recovered mixture fraction must satisfy $0\leq\overline{Z}\leq1$. Each trial pressure must additionally produce positive density; otherwise, the trial state is discarded.

When the mixed normal velocity is small, the pressure root may lie within a narrow interval close to $p_{\mathrm{hi}}$. A uniformly spaced pressure scan can therefore miss the corresponding sign change in $\mathcal{R}_p$. To increase the resolution near the upper boundary, the initial pressure samples are distributed as
\begin{equation}
    p_i = p_{\mathrm{hi}}-\left(p_{\mathrm{hi}}-p_{\mathrm{lo}}\right)\left(\frac{i}{N_p-1}\right)^{\beta},
    \qquad
    i=0,\ldots,N_p-1,
    \qquad
    \beta>1,
\label{eq:clustered_pressure_sampling}
\end{equation}
where $N_p$ is the number of pressure samples. The present calculations use $\beta=2$, which clusters the samples near $p_{\mathrm{hi}}$ while retaining coverage of the complete interval; the impact of $\beta$ on the solution will be further discussed in Section~\ref{subsec:u approx 0}.

For each sampled pressure, the temporary velocity, density, required static enthalpy, temperature, and pressure residual are evaluated using the recovery procedure described above. Two neighbouring admissible samples define a root bracket when
\begin{equation}
    \mathcal{R}_p(p_i) \mathcal{R}_p(p_{i+1})<0.
\label{eq:pressure_bracket}
\end{equation}
If a sampled point already satisfies the prescribed pressure tolerance, it is accepted directly. Otherwise, the root within each sign-changing interval is refined using bisection. This bracketed procedure keeps all pressure iterates inside the admissible interval and avoids differentiating the embedded temperature inversion.

The boundary-clustered search specifically addresses the limit $u_n\rightarrow0$ while $f_m$ remains nonzero. The distinct limit $f_m\rightarrow0$ makes $\overline{Z}=f_Z/f_m$ and $\overline{h}_t=f_h/f_m$ ill-conditioned and is therefore outside the pressure-based reconstruction considered here.

\FloatBarrier

\section{Computational framework}
\label{sec:computational_framework}

The simulations are performed using an in-house three-dimensional unstructured finite-volume solver developed by Wang et al.~\cite{wang2016virtual,wang2023gpu}. The steady Reynolds-averaged Navier-Stokes (RANS) equations are discretised using a cell-centred finite-volume method. Flow-variable gradients are reconstructed using a weighted least-squares method. Convective fluxes are evaluated using the Roe's approximate Riemann solver~\cite{roe1981approximate}, together with second-order MUSCL reconstruction and the van Albada limiter~\cite{van1982comparative}, whereas viscous fluxes are discretised using central differences. The Wilcox $k$--$\omega$ model~\cite{wilcox1998turbulence} is employed for turbulence closure. A matrix-free implicit scheme based on flux linearisation is used for pseudo-time advancement. The spatial discretisation is nominally second-order accurate. The solver has previously been validated for a range of aero-engine intake and compressor flows~\cite{carnevale2016lip,wang2018numerical,wang2018simulation}, supporting its application to the present simulations.

\section{Results and discussion}
\label{sec:results}
This section evaluates the proposed reacting-mixture mixing-plane method through three test cases. First, an idealised interface test is performed using prescribed upstream states. This case examines the influence of fuel type and mixture fraction distribution on the recovered thermodynamic state, and a thorough investigation of recovery at $\bar{u}_n \approx 0$ is also conducted. Second, the Darmstadt transonic compressor is used to demonstrate that for a perfect gas, the proposed formulation reduces to the conventional mixing-plane method, thereby providing a consistency check against established non-reacting practice. Finally, an integrated KJ66 micro-turbine engine simulation is conducted to demonstrate the proposed method in a whole-engine multiphysics configuration and to assess key performance quantities, including mass flow rate, pressure ratio, and thrust, through comparison with available experimental data.

\subsection{1-D theoretical test case}
\label{subsec:1D test}
\begin{figure}[htbp]
\centering
\includegraphics[width=0.5\columnwidth]{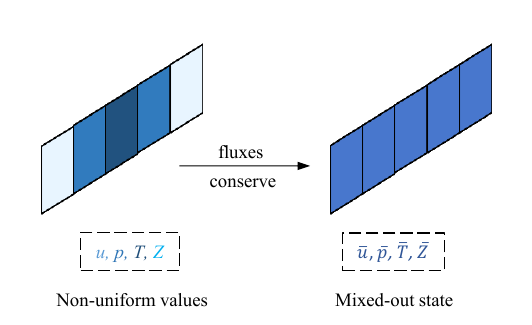}
\caption{Schematic of the one-dimensional consistency test, in which non-uniform flow and mixture fraction profiles are reconstructed as a mixed-out state with identical conservative fluxes.}
\label{fig:1d_view}
\end{figure}

The theoretical test isolates the mixed-out state recovery from grid resolution, turbulence closure, and three-dimensional blade-row effects. A one-dimensional interface coordinate $y\in[0,1]$ is prescribed with non-uniform primitive fields $u(y)$, $p(y)$, $T(y)$, and mixture fraction $Z(y)$. These fields are first converted to local thermochemical states and then integrated to obtain the area-averaged mass flux, momentum flux, total-enthalpy flux, and scalar flux. The proposed algorithm receives only these four fluxes and reconstructs a single mixed-out state $(\bar{\rho},\bar{u},\bar{p},\bar{T},\bar{Z})$ that satisfies the same flux balances and the mixture equation of state. A schematic of the test is presented in Figure~\ref{fig:1d_view}.

The test is therefore a theoretical consistency test rather than a direct model of a realistic interface. It provides a controlled environment for checking whether the nonlinear recovery preserves the imposed conservative quantities and how the pressure/temperature iterations respond to mixture distribution, fuel thermochemistry, and pressure-root admissibility. The reference case uses H$_2$ with a smooth Gaussian mixture fraction profile, $Z(y)=\exp[-(y-0.5)^2/(2\times0.12^2)]$. Table~\ref{tab:quasi1d_flux_reference} compares the imposed fluxes with the reconstructed mixed-out fluxes for this case.


\begin{table}[htbp]
\centering
\footnotesize
\caption{Flux conservation in the reference quasi-1D H$_2$ Gaussian case.}
\label{tab:quasi1d_flux_reference}
\begin{tabular}{lcccc}
\toprule
Flux & Relative difference \\
\midrule
  Mass & $2.18\times10^{-16}$ \\
  Momentum & $1.77\times10^{-15}$ \\
  Energy & $3.63\times10^{-13}$ \\
  Mixture & $2.96\times10^{-16}$ \\
\bottomrule
\end{tabular}
\end{table}

The reference result confirms that the recovered mixed-out state preserves the conservative fluxes to round-off accuracy. This conservation check is used as the baseline before considering profile shape, fuel kind, and pressure-root location.

\subsubsection{Impact of mixture profiles}
Consider the mixture fraction distribution changes for increasing the non-uniformity while keeping the fuel fixed as H$_2$. Five profile families are considered: Gaussian, top-hat, hyperbolic tangent, sinusoidal, and multi-Gaussian. The profiles are designed to cover smooth localised non-uniformity, discontinuous-like scalar jumps, monotonic transition, oscillatory variation, and multiple scalar peaks. All profiles are bounded by $0\le Z\le1.0$. Table~\ref{tab:h2_profile_parameters} gives the parameters of the profiles used.

\begin{table}[htbp]
  \centering
  \footnotesize
  \caption{Parameters of the distinct mixture fraction profiles.}
  \label{tab:h2_profile_parameters}
  \begin{tabular}{ll}
  \toprule
  Profile & Parameters \\
  \midrule
  Gaussian & $a=1$ centred at $y=0.5$, $\sigma=0.12$ \\
  Top-hat & $Z=1$ within $|y-0.5|\le0.16$, $Z=0$ elsewhere \\
  Tanh & $a=1$ centred at $y=0.5$, transition width $\sigma=0.04$ \\
  Sinusoidal & $Z_{ave}=0.5$, $a=0.5$, wavenumber $k=2$ \\
  Multi-Gaussian & Two peaks, $k=2$, $\sigma=0.08$, $a=2.0$ \\
  \bottomrule
  \end{tabular}
\end{table}

\begin{figure}[htbp]
\centering
\includegraphics[width=0.6\columnwidth]{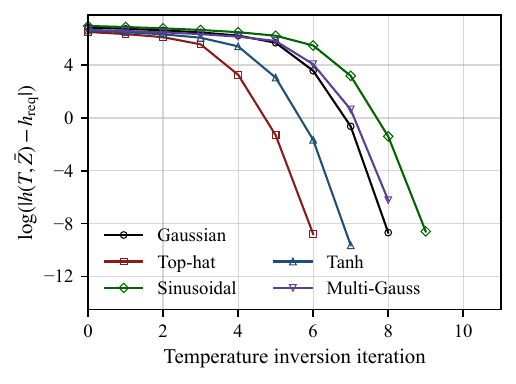}
\caption{Temperature-inversion residual histories during the first pressure iteration for five mixture-fraction profiles in the 1-D test.}
\label{fig:quasi1d_h2_profile_enthalpy_convergence}
\end{figure}

\begin{figure}[htbp]
\centering
\includegraphics[width=0.6\columnwidth]{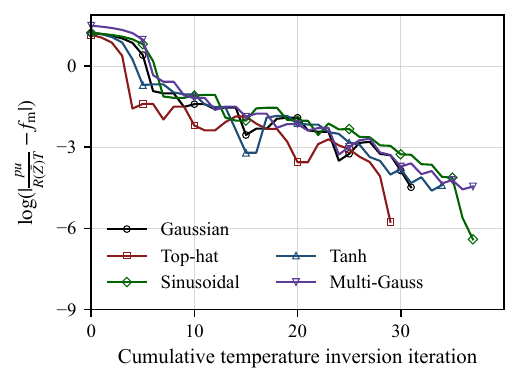}
\caption{Evolution of the EOS-based mass-flux mismatch during mixed-state recovery for five mixture-fraction profiles in the 1-D test.}
\label{fig:quasi1d_h2_profile_mass_convergence}
\end{figure}

Figure~\ref{fig:quasi1d_h2_profile_enthalpy_convergence} reports the temperature inversion enthalpy residual during the first outer pressure iteration, thereby isolating the initial difficulty of the temperature inversion before the pressure bracketing and bisection processes dominate. Figure~\ref{fig:quasi1d_h2_profile_mass_convergence} further shows the evolution of an EOS-based mass-flux mismatch over the full mixed-state recovery process. At each iteration, the velocity is determined from the prescribed mass and momentum fluxes, whereas the density associated with the current pressure and temperature is evaluated through the equation of state. The resulting mismatch therefore measures the inconsistency between the mass flux implied by the current thermodynamic state and the prescribed mass flux. It approaches zero as the recovered state satisfies the thermodynamic closure.

All five H$_2$ profile cases converged. The total number of inner temperature-inversion iterations was 31 for the Gaussian case, 29 for the top-hat case, 34 for the tanh case, 36 for the sinusoidal case and the multi-Gaussian case. The top-hat profile produced the largest scalar-gradient metric while it produced the smallest iteration count. These results indicate that mixture-profile shape changes the transient residual path, but the recovery is not controlled by scalar-gradient magnitude alone.

\subsubsection{Impact of fuel types}
Changing the fuel can lead to significant variations in thermodynamic properties even when the mixture-fraction profile is fixed. The tested fuels are H$_2$, CH$_4$, C$_7$H$_{16}$, C$_{10}$H$_{22}$, and C$_{12}$H$_{26}$. To compare them on a physically scaled basis, the same baseline Gaussian profile $Z(y)=\exp[-(y-0.5)^2/(2\times0.12^2)]$ is used.

An indicator is introduced to describe the thermochemical difficulty of the fuel cases. Since the relation $h(T,\bar{Z})$ is nonlinear and implicit through the thermochemical model, $T$ is recovered via a Newton iteration of Eqs.~(\ref{eq:newton_t}) and (\ref{eq:res_t}), defining the error $e^{l}=T^{l}-T$.
To quantify the local convergence behaviour, $\mathcal{R}_T$ about the solution $T$ is expanded as
\begin{equation}
    \mathcal{R}_T(T^{l})=\mathcal{R}_T(T)+\mathcal{R}_T'(T)(T^{l}-T)+\frac{1}{2}\mathcal{R}_T''(T)(T^{l}-T)^2.
\end{equation}
Using $\mathcal{R}_T(T)=0$, the Newton update yields the quadratic error estimate
\begin{equation}
    |e^{l+1}|\le \frac{1}{2}\left|\frac{\mathcal{R}_T''(T)}{\mathcal{R}_T'(T)}\right|\,|e^{l}|^2
    =\frac{1}{2}\left|\frac{\partial c_p/\partial T}{c_p}\right|\,|e^{l}|^2.
\end{equation}
Accordingly, a physically interpretable thermochemical stiffness indicator can be introduced as\par
\begin{equation}
    \chi_{\mathrm{T}}\sim \left|\frac{1}{c_p}\frac{\partial c_p}{\partial T}\right|.
\end{equation}
This quantity also characterises the thermodynamic stiffness: in high-temperature regions, or for the fuels where $c_p$ varies more rapidly with $T$, the stiffness increases and the inner Newton iteration convergence may be affected.

\begin{figure}[htbp]
\centering
\includegraphics[width=0.6\columnwidth]{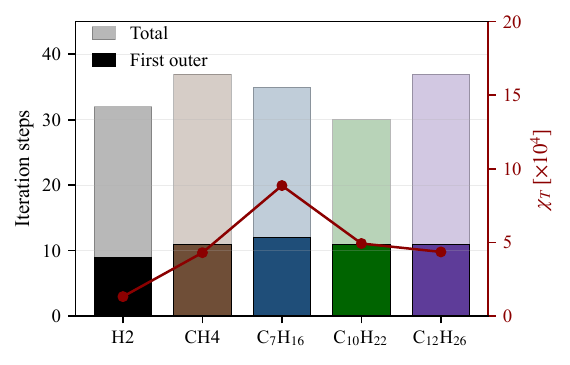}
\caption{Fuel-type dependence of recovery cost and thermochemical stiffness for a baseline Gaussian profile.}
\label{fig:quasi1d_fuel_kind_iterations}
\end{figure}

The results in Figure~\ref{fig:quasi1d_fuel_kind_iterations} suggest that fuels with larger values of $\chi_T$ generally require more temperature inversion steps during the first outer iteration, indicating that the nonlinear enthalpy-temperature relation becomes more demanding when the thermochemical properties vary more rapidly with temperature. However, this trend does not uniquely determine the total number of recovery iterations. The overall convergence cost is also governed by the pressure-root localisation process, the coupling among the conserved flux constraints, the distance between the initial estimate and the admissible thermodynamic state, and possible boundary-root or inadmissible-root situations. Therefore, the thermodynamic stiffness should be interpreted as a local indicator of the temperature inversion difficulty rather than a complete predictor of the global convergence efficiency.

\subsubsection{Treatment of $\bar{u}_n \approx 0$}
\label{subsec:u approx 0}
As discussed in Section~\ref{subsec:admissible_pressure_search}, the pressure-root search requires special attention when the recovered normal velocity is close to zero. For the one-dimensional recovery problem, the mixed velocity and density can be written as
\begin{equation}
    \bar{u}=\frac{f_{Mx} - \bar{p}}{f_m},\qquad
    \bar{\rho}=\frac{f_m^2}{f_{Mx} - \bar{p}}.
\end{equation}
The upper admissible pressure boundary is therefore $p_{\max}=f_{Mx}$. When the normal velocity approaches zero, the admissible pressure root may lie extremely close to this upper boundary. To quantify this behaviour, a relative root-position parameter is introduced as
\begin{equation}
  \chi_p = \frac{f_{Mx}-\bar{p}}{f_{Mx}}.
\end{equation}
The parameter \(\chi_p\) measures the fraction of the streamwise momentum flux that remains as dynamic pressure. Thus, $\chi_p \to 0$ corresponds to the low Mach limit, in which the recovered pressure approaches the upper bound. In this limit, a uniformly sampled pressure scan may fail to resolve the narrow interval between the physical root and $p_{\max}$. For this reason, the bracket search employs the power-law pressure sampling defined in Eq.~(\ref{eq:clustered_pressure_sampling}), which clusters trial samples near the upper admissible boundary.

\begin{figure}[htbp]
\centering
\includegraphics[width=0.6\columnwidth]{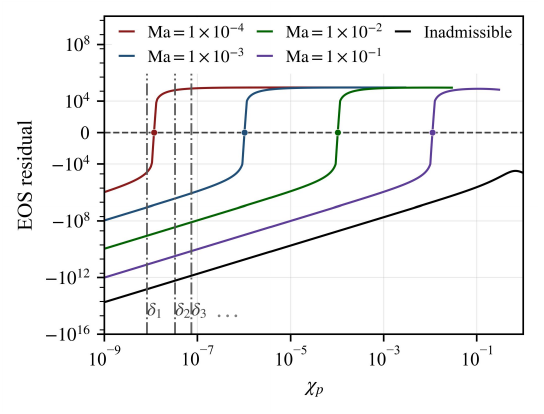}
\caption{EOS pressure-residual behaviour for an admissible root close to $p_{\max}$ and an inadmissible state. Grey lines denote the sampled pressure locations near the upper admissible boundary.}
\label{fig:quasi1d_boundary_pressure_residual}
\end{figure}

The limiting behaviour as $\bar{u_n} \rightarrow 0$ is examined by progressively reducing the imposed velocity scale while retaining the same baseline $\mathrm{H}_2$ Gaussian profile. As shown in Figure~\ref{fig:quasi1d_boundary_pressure_residual}, decreasing the Mach number shifts the zero crossing of the EOS pressure residual towards smaller values of $\chi_p$, indicating that the physical root moves closer to $p_{\max}$. The grey vertical lines denote the first several nonzero normalised pressure distances generated by the quadratic power-law sampling strategy. In this test, $\beta=2$ and $N_p=10000$ are used. For the lowest velocity scale, $\mathrm{Ma}=1\times10^{-4}$, the sign change still exists and is captured within the first few sampled intervals adjacent to $p_{\max}$. As the velocity scale increases, the zero crossing moves farther away from the upper boundary and can be detected more readily by the bracketing procedure.

The inadmissible case shown in Figure~\ref{fig:quasi1d_boundary_pressure_residual} is qualitatively different from the case of an admissible root located close to $p_{\max}$. It is generated from a synthetic $\mathrm{H}_2$ multi-Gaussian scalar profile with $Z_{\mathrm{base}}=0$, amplitude $Z_{\mathrm{amp}}=1$, $\sigma=0.06$, and two scalar peaks under the same imposed velocity, pressure, and temperature fields. In this case, the EOS pressure residual does not exhibit a sign change within the sampled pressure interval, indicating that no admissible mixed-out state is found for the prescribed synthetic profile and thermochemical closure. Therefore, the convergence risk in this test is associated with two distinct issues: the existence of a physically admissible root and the ability of the pressure sampling strategy to bracket a root close to $p_{\max}$.

To evaluate the effect of the sampling parameters on bracket detection, Table~\ref{tab:quasi1d_bracket} summarises the results obtained from four parameter configurations. Here, $\delta_1$ denotes the first nonzero normalised distance from the upper pressure boundary, and $\Delta\delta_{\mathrm{root}}$ denotes the width of the detected root bracket in the same normalised coordinate. For the second configuration, the first valid sampled pressure is already located beyond the root under the low Mach condition; consequently, the sign-changing interval is missed. The other three configurations successfully identify the root bracket. The first and the third configurations produce smaller bracket widths and therefore require fewer subsequent bisection iterations. In the following computations, the $\beta=2, N_p=10000$ configuration is adopted. It should be noted that these parameters are not strictly fixed and may be adjusted according to the relative levels of dynamic and static pressure. 
For cases with $Ma < 10^{-4}$, further refinement of the pressure sampling is not pursued. At such low velocity levels, the corresponding mass flux is extremely small, and strict enforcement of its conservation may introduce numerical difficulties without producing a meaningful influence on the overall solution. In the present implementation, the conservation constraint associated with this negligible mass flux is therefore relaxed.

\begin{table}[htbp]
\centering
\footnotesize
\caption{Sensitivity of bracket detection to power-law pressure-sampling parameters.}
\label{tab:quasi1d_bracket}
\begin{tabular}{lcccc}
\toprule
$\beta$ & $N_p$ & $\delta_1$ & $\Delta\delta_\mathrm{root}$ & Bisection steps \\
\midrule
  2 & $10000$ & $1.000\times10^{-8}$ & $3.00\times10^{-8}$ & 16 \\
  2 & $1000$  & $1.002\times10^{-6}$ & -                   & -  \\
  3 & $1000$  & $1.003\times10^{-9}$ & $1.91\times10^{-8}$ & 16 \\
  4 & $100$   & $1.041\times10^{-8}$ & $1.56\times10^{-7}$ & 18 \\
\bottomrule
\end{tabular}
\end{table}

\begin{figure}[htbp]
\centering
\includegraphics[width=0.6\columnwidth]{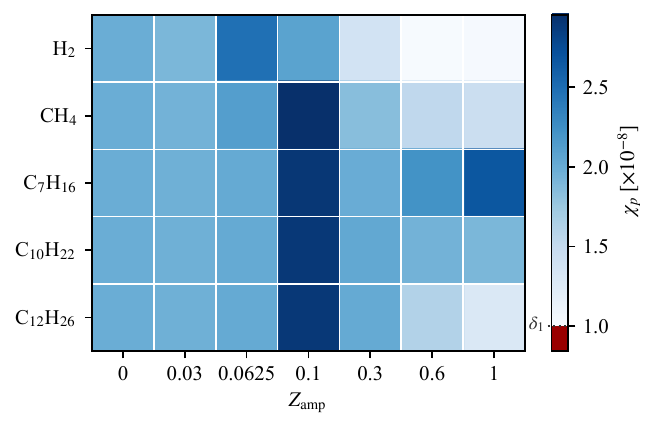}
\caption{Effects of fuel type and mixture-fraction profile on the pressure-root location at $\mathrm{Ma}=1 \times 10^{-4}$.}
\label{fig:quasi1d_boundary_fuel}
\end{figure}

Figure~\ref{fig:quasi1d_boundary_fuel} further examines the effects of fuel chemistry and mixture fraction distribution on the pressure-root location in the low Mach-number limit. The velocity scale is fixed at $\mathrm{Ma}=1\times10^{-4}$, and seven representative Gaussian scalar profiles with $Z_{\mathrm{amp}}=(0, 0.03, 0.0625, 0.1, 0.3, 0.6, 1.0)$ are evaluated for $\mathrm{H_2}$, $\mathrm{CH_4}$, $\mathrm{C_7H_{16}}$, $\mathrm{C_{10}H_{22}}$, and $\mathrm{C_{12}H_{26}}$. The colormap denotes the normalised pressure-root position, $\chi_p$. Roots located at smaller \(\chi_p\) than the first sampled pressure point would be highlighted in dark red, indicating cases in which the bracketing strategy would fail to localise the sign-changing interval. In the present tests, however, the first sampled point lies at smaller $\chi_p$ than all admissible roots, and therefore all root intervals are successfully bracketed. The results show that, under this low Mach condition, the admissible pressure root is confined to an $\mathcal{O}(10^{-8})$ neighbourhood of $f_{Mx}$, while its exact location depends noticeably on both fuel type and $Z_\mathrm{amp}$. In particular, several high $Z_\mathrm{amp}$ hydrogen cases place the root close to the first sampling point, demonstrating that the pressure scan must retain sufficient resolution near $f_{Mx}-\bar{p}\to 0$ to avoid missing admissible roots.

\FloatBarrier

\subsection{Darmstadt transonic compressor}
A generalised reacting-mixture mixing-plane formulation should recover the conventional flux-conservative mixing-plane method as a limiting case when the working fluid is modelled as perfect-gas air. Verification of this consistency is important because it demonstrates that the proposed approach can be applied uniformly to both reacting and non-reacting regions within a single computational framework. The Darmstadt transonic compressor is therefore selected as a three-dimensional non-reacting consistency test for the proposed reacting-mixture mixing-plane formulation. In this case, the scalar is prescribed as $Z=0$, and the reconstructed composition corresponds to standard air. Under this condition, the scalar flux is identically zero and the thermochemical closure reduces to a fixed-composition air model. The proposed formulation should then recover the same mixed-out pressure, density, velocity components, temperature, Mach number, and total enthalpy as the conventional conservative mixing-plane treatment. 

The Darmstadt compressor is simulated as a verification test case using the TUDa-GLR-OpenStage configuration, an open transonic axial-compressor dataset developed at TU Darmstadt for numerical-model assessment and solver validation. The compressor stage consists of a 16-blade blisk rotor, a 29-vane stator, and a 5-vane outlet guide vane (OGV). The rotor tip clearance is approximately $0.8\%$ of the blade span. The design rotational speed is about $20{,}000~\mathrm{rpm}$, with a design mass flow rate of approximately $16~\mathrm{kg/s}$ and a total pressure ratio of about $1.5$ \cite{klausmann2022tuda}. Further details of the investigated stage are summarised in Table~\ref{tab:tech_data_tud}.

\begin{table}[htbp]
\footnotesize
\setlength{\tabcolsep}{5pt}
\caption{Technical data of the Darmstadt compressor}
\label{tab:tech_data_tud}
\centering{
\begin{tabular}{!{\hspace*{0.25cm}} 
                >{\raggedright\hangindent=1em}p{5.5cm}
                >{\raggedright\arraybackslash}p{1.2cm} 
                !{\hspace*{0.25cm}}}
\toprule
Parameter & Value \\
\midrule
Maximum power [kW] & 800 \\
Maximum torque [Nm] & 350 \\
Design rotational speed [rpm] & 20,000 \\
Design mass flow rate [kg/s] & 16 \\
Casing radius [mm] & 189.2 \\
Rotor tip clearance [mm] & 0.8 \\
Number of rotor blades [-] & 16 \\
Number of stator vanes [-] & 29 \\
Number of outlet guide vanes [-] & 5 \\
\bottomrule
\end{tabular}
}
\end{table}

\begin{figure}[htbp]
\centering
\includegraphics[width=0.6\columnwidth]{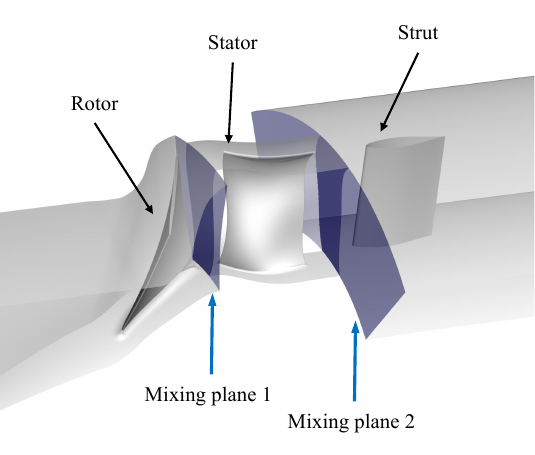}
\caption{Computational domain and locations of the two mixing planes (MP1 and MP2) in the Darmstadt compressor.}
\label{fig:tuda_geometry}
\end{figure}

\begin{figure*}[htbp]
\centering
\includegraphics[width=0.5\textwidth]{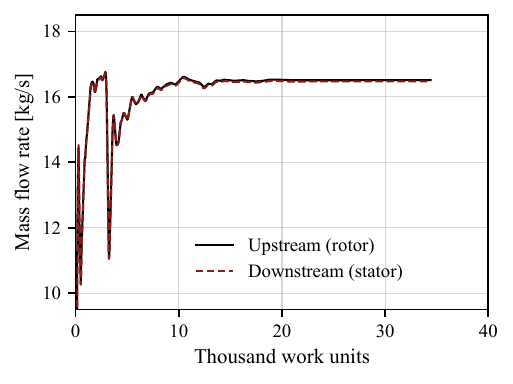}
\caption{Mass flow convergence history at the peak-efficiency operating point of the Darmstadt compressor at $100\%$ design speed.}
\label{fig:tud_convergence}
\end{figure*}

The simulations are performed using a single-passage flow domain with circumferentially periodic boundary conditions, and focus primarily on the $100\%$ design-speed condition. The computational domain and mixing-plane locations are illustrated in Figure~\ref{fig:tuda_geometry}. The measured total pressure and total temperature profiles are imposed at the inlet, and the inlet flow direction is specified as axial. Mixing-plane boundary conditions are applied at both the rotor--stator and stator--OGV interfaces. At the outlet, a radial-equilibrium back-pressure condition is prescribed. The convergence history of the peak-efficiency (PE) point is shown in Figure~\ref{fig:tud_convergence}. Convergence was achieved after approximately 32,000 iterations at the PE point, all other conditions were likewise computed to convergence, and the last converged operating point is regarded as the numerical stall point.

\begin{figure}[htbp]
\centering
\includegraphics[width=0.5\textwidth]{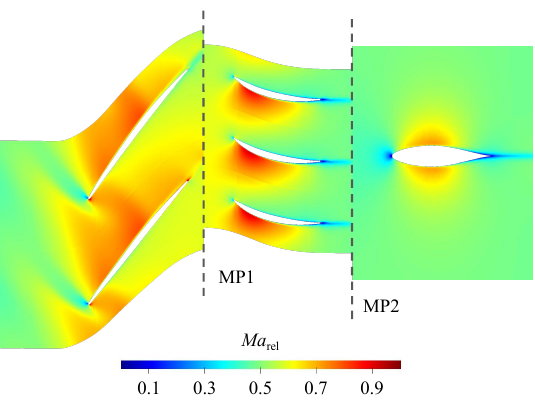}
\caption{Relative Mach-number contours at $50\%$ span for the peak-efficiency operating point at $100\%$ design speed, with MP1 and MP2 indicated.}
\label{fig:tuda_marel}
\end{figure}

Figure~\ref{fig:tuda_marel} presents the relative Mach number distribution at $50\%$ span at the PE operating point. The rotor accelerates the incoming flow along the suction surface, producing a highly non-uniform velocity field at the rotor exit. This non-uniform flow is transferred across the rotor-stator mixing plane (MP1) and subsequently diffused within the stator passage. At the stator exit (MP2), the flow becomes considerably more uniform, illustrating the flow-averaging process that occurs across the blade-row interfaces.

\begin{figure}[htbp]
\centering
\includegraphics[width=0.46\textwidth]{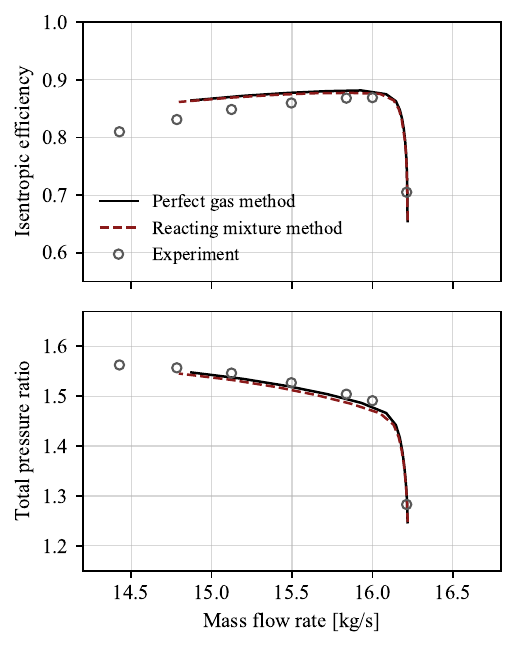}
\caption{Compressor characteristics at $100\%$ design speed: isentropic efficiency and total pressure ratio predicted by the perfect gas and reacting-mixture methods compared with experimental data.}
\label{fig:tud_performance}
\end{figure}

To verify the general applicability of the proposed method, two sets of simulations are performed under the same non-reacting flow condition. The first set employs the conventional flux-conservation mixing-plane method, hereafter referred to as the perfect-gas (PG) method, whereas the second set employs the proposed reacting-mixture (RM) method. The stage performance characteristics at the design speed are presented in Figure~\ref{fig:tud_performance}. Good agreement with the experimental data was obtained, with PG method and RM method producing nearly identical results. The isentropic efficiency reaches its maximum at a mass flow rate of approximately $16~\mathrm{kg/s}$. This operating condition is therefore selected as the PE point and is used as the representative case for the subsequent analysis.

\begin{figure*}[htbp]
\centering
\includegraphics[width=0.95\textwidth]{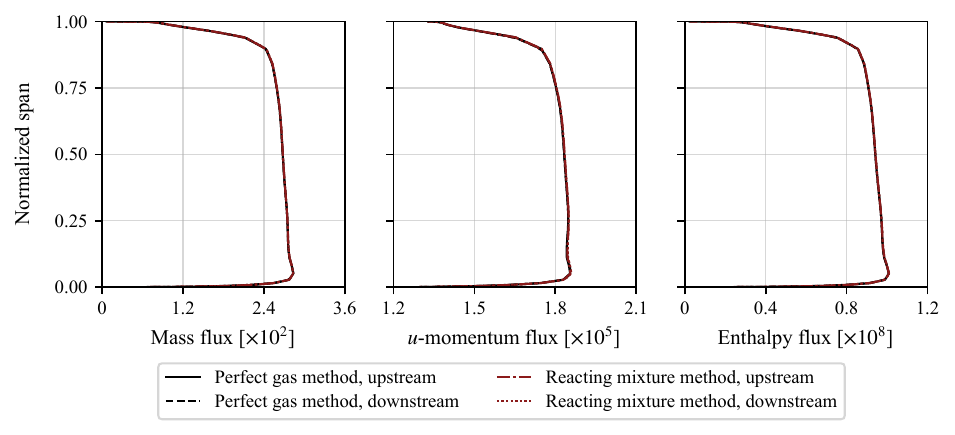}
\caption{Upstream and downstream spanwise distributions of mass, streamwise-momentum, and total-enthalpy fluxes at the rotor–stator mixing plane, comparing the perfect gas and reacting-mixture methods.}
\label{fig:tud_radial_1}
\end{figure*}

The radial distribution of fluxes at the rotor--stator interface is presented in Figure~\ref{fig:tud_radial_1}. Both the PG and RM methods employ the same upstream interface fluxes, and the downstream fluxes imposed from the reconstructed mixed-out state are nearly identical. This consistency verifies that the proposed method retains the same flux-conservation behaviour as the conventional mixing-plane method.

\begin{figure}[htbp]
\centering
\includegraphics[width=0.8\columnwidth]{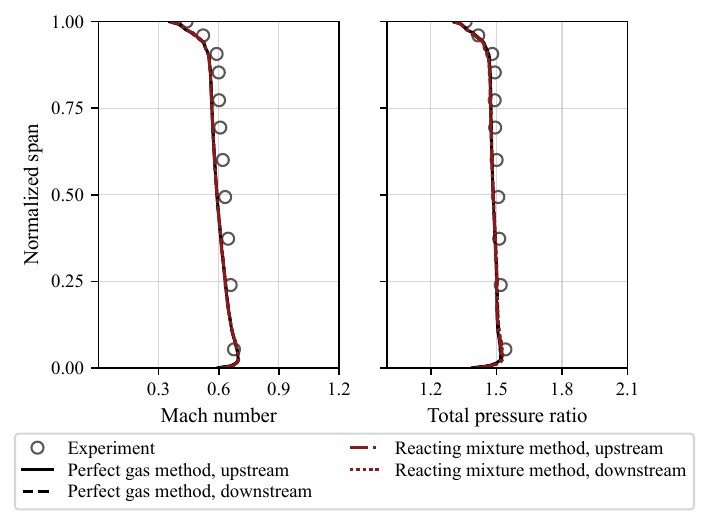}
\caption{Spanwise Mach-number and total pressure-ratio profiles at the rotor–stator mixing plane, comparing experimental data with the perfect gas and reacting-mixture methods.}
\label{fig:tud_radial_2}
\end{figure}

\begin{figure}[htbp]
\centering
\includegraphics[width=0.8\columnwidth]{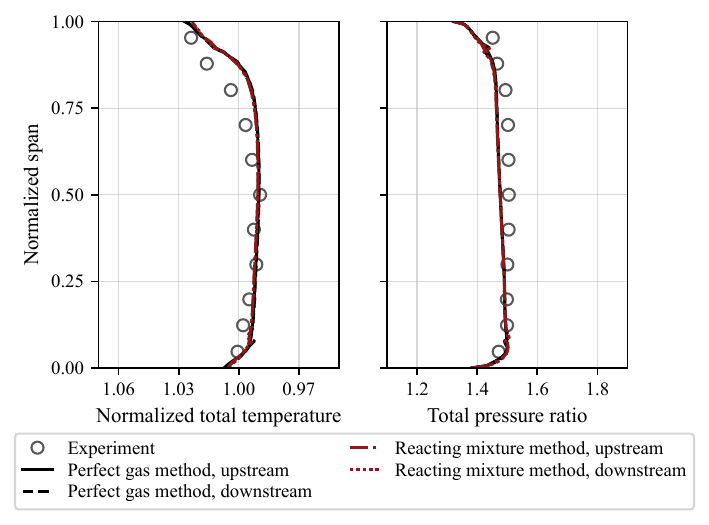}
\caption{Spanwise profiles of normalised total temperature and total pressure ratio across the stage-exit mixing plane, comparing experimental data with the perfect gas and reacting-mixture methods.}
\label{fig:tud_radial_3}
\end{figure}

Figure~\ref{fig:tud_radial_2} illustrates the radial profiles at the first rotor–stator interface. Similarly, the profiles at the second mixing plane, which is located between the stage exit and the OGV inlet, are compared in Figure~\ref{fig:tud_radial_3}. These profiles provide a detailed assessment of the spanwise distributions of the total pressure ratio, total temperature, and Mach number and further demonstrate consistency with the experimental measurements.

\begin{figure}[htbp]
\centering
\includegraphics[width=0.5\columnwidth]{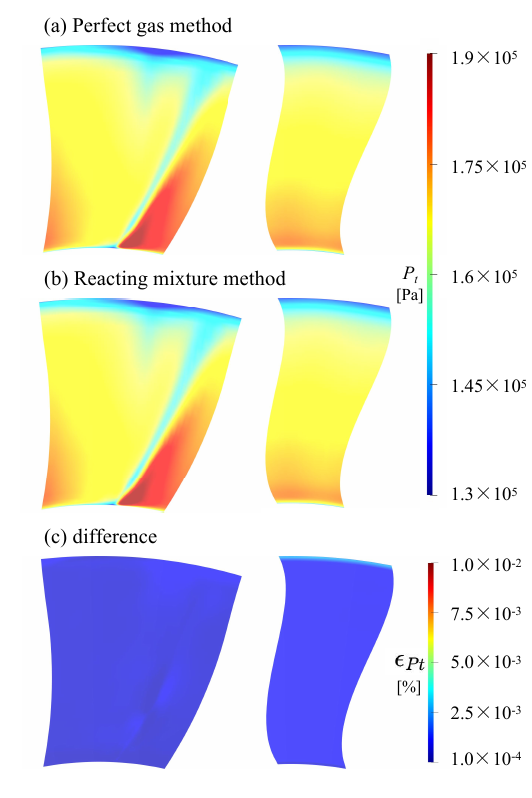}
\caption{Total pressure contours at the rotor exit (left) and stator inlet (right) for the non-reacting Darmstadt case: (a) perfect gas method, (b) reacting-mixture method with $Z=0$, and (c) relative difference.}
\label{fig:tud_pt}
\end{figure}

Figure~\ref{fig:tud_pt} compares the flow contours at rotor--stator interface obtained using the classical method and the proposed formulation with $Z=0$. Because both methods enforce identical mass, momentum, and total enthalpy flux constraints under fixed-composition thermodynamics, their solutions are expected to be virtually identical. The good agreement between the two methods demonstrates that the reacting-mixture formulation introduces no artificial thermochemical corrections in the non-reacting flow, while successfully retaining the conservative properties of the established approach.

\FloatBarrier

\subsection{Whole-engine simulation of the KJ66 micro turbine engine}

\begin{figure}[htbp]
\centering
\includegraphics[width=0.6\columnwidth]{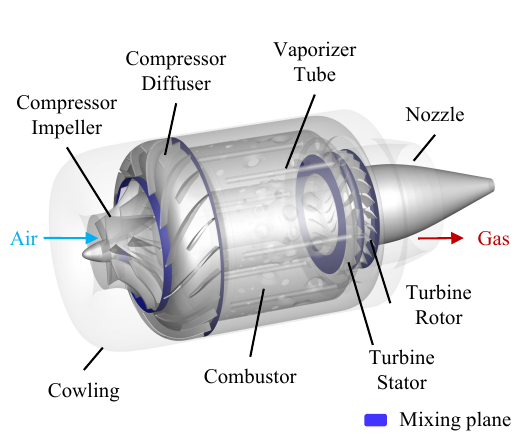}
\caption{Schematic of the KJ66 micro turbine engine, showing the major components, flow direction, and five mixing plane locations.}
\label{fig:kj66_geo}
\end{figure}

To demonstrate the aforementioned properties of the new method, the KJ66 micro gas turbine engine (MTE), an axial-centrifugal turbomachinery configuration with publicly available experimental data, has been used as a test case. Although small in scale, the KJ66 is a representative turbojet engine containing all essential subsystems, including the nacelle, compressor, combustor, turbine, and nozzle. The illustration of the three-dimensional geometry is shown in Figure~\ref{fig:kj66_geo}. Five mixing planes between the components are highlighted. The first two interfaces transfer non-reacting air, whereas the three interfaces downstream of the combustor transfer reacting mixtures. This configuration therefore provides a suitable environment for assessing the proposed algorithm across both non-reacting and reacting-mixture conditions.

\begin{figure}[ht]
\centering
\includegraphics[width=0.95\columnwidth]{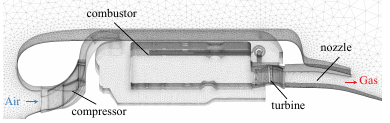}
\caption{Meridional computational mesh of the KJ66 micro turbine engine.}
\label{fig:kj66_mesh}
\end{figure}

The design parameters and experimental data of the KJ66 MTE are summarised in Table~\ref{tab:tech_data_kj66}. These data include the blade and vane counts of the compressor and turbine, the number of vaporizer tubes in the combustor, and the measured performance quantities at a rotational speed of 100,000 rpm, including thrust, fuel consumption, and mass flow rate. All reference data are taken from the literature \cite{Schreckling2005}. The whole-engine meshes were generated using Pointwise. For steady-state simulations, the circumferential periodicity of each component was utilised to reduce the computational cost by modelling only a single-sector, single-passage domain. The near-wall grid spacing was selected to maintain $y^+ \approx 1$ on solid surfaces. Additional local refinements were applied in regions of rotor tip clearance, around the combustor vaporizer-tube walls, and near the fuel-injection ports. The final computational mesh contains approximately 6.03 million cells. The mesh distribution is shown in Figure~\ref{fig:kj66_mesh}.

\begin{table}[htbp]
\footnotesize
\setlength{\tabcolsep}{5pt}
\caption{Technical data of the KJ66 MTE}
\label{tab:tech_data_kj66}
\centering{
\begin{tabular}{!{\hspace*{0.25cm}} 
                >{\raggedright\hangindent=1em}p{5.5cm}
                >{\raggedright\arraybackslash}p{1.2cm} 
                !{\hspace*{0.25cm}}}
\toprule
Parameter & Value \\
\midrule
Weight [g] & 930 \\
Length [mm] & 230 \\
Max. diameter [mm] & 112 \\
Compressor impeller blade count [-] & 6 \\
Compressor diffuser vane count [-] & 12 \\
Turbine stator vane count [-] & 18 \\
Turbine rotor blade count [-] & 24 \\
Vaporizer tube count [-] & 6 \\
Max. design speed [rpm] & 120,000 \\
\midrule
Data at 100,000 rpm: & \\
\midrule
Thrust [N] & 52 \\
Exhaust gas temperature [K] & 853 \\
Pressure ratio [-] & 1.88 \\
Outflow velocity [m/s] & 365 \\
Kerosene consumption [ml/min] & 300 \\
Air mass flow [kg/s] & 0.192 \\
\bottomrule
\end{tabular}
}
\end{table}

\begin{figure}[H]
\centering
\includegraphics[width=0.5\textwidth]{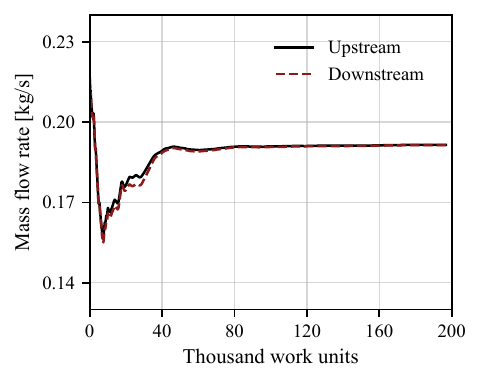}
\caption{Upstream and downstream mass flow histories at the combustor–turbine mixing plane during convergence of the KJ66 simulation at $1\times10^5$ rpm.}
\label{fig:kj66_convergence}
\end{figure}

The whole-engine multiphysics steady flow simulation is performed using an in-house turbomachinery simulation platform \cite{wang2023gpu}. The simulation requires boundary conditions to be specified at the impeller inlet, nozzle far-field boundary and the combustor fuel outlet at a rotational speed of 100,000 rpm. To be more specific, total pressure, temperature and flow angles are specified at the impeller inlet, static back pressure is specified at the far-field and a mass flow injection boundary is specified at the fuel outlet. All solid walls are modelled as viscous, adiabatic walls. The proposed method is used to couple the engine components. The combustion process is modelled using a flamelet approach with gaseous n-decane ($\mathrm{C}_{10}\mathrm{H}_{22}$) as the surrogate fuel. A GPU server that consists of eight NVIDIA RTX 5880 Ada GPUs is used to perform the computation. The convergence history of mass flow rate is shown in Figure~\ref{fig:kj66_convergence}. The steady computation converged after 198,000 iterations for the whole-engine simulation.

\begin{figure}[ht]
\centering
\includegraphics[width=1.0\columnwidth]{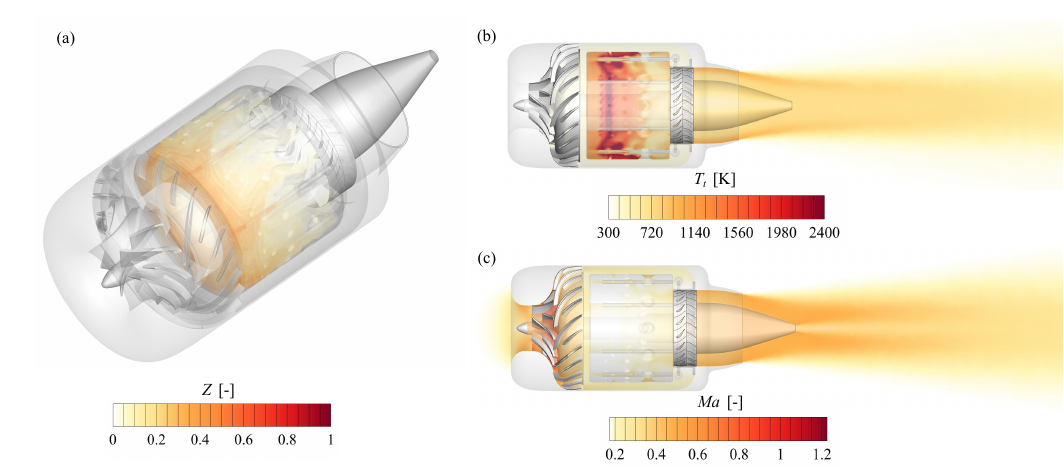}
\caption{Meridional distributions of (a) mixture fraction, (b) total temperature, and (c) Mach number in the KJ66 MTE at $1\times10^5$ rpm.}
\label{fig:kj66_flowfield1}
\end{figure}

The meridional plane distribution of mixture fraction, total temperature and Mach number computed at a rotational speed of 100,000 rpm is shown in Figure~\ref{fig:kj66_flowfield1}. The computed flow field captures the expected operating sequence of the KJ66 MTE. The incoming air is compressed and accelerated through the centrifugal compressor impeller, followed by further static-pressure recovery and flow deceleration in the diffuser. Downstream of the compressor, the injected fuel is introduced into the combustor outer annulus, where it mixes with the compressor discharge air transferred through the compressor--combustor mixing plane. The mixture fraction distribution indicates that the fuel–air mixture remains highly non-uniform near the injection region, whereas stronger mixing is achieved as the flow is entrained into the combustor recirculation zone. This recirculating flow promotes flame stabilisation and enhances fuel–air mixing, leading to a high-temperature combustion region and associated volumetric expansion. The resulting hot reacting mixture then passes through the combustor--turbine mixing plane which represents the most relevant application scenario for the proposed method, because the transferred flow contains strong thermal and compositional non-uniformities generated by combustion. After entering the turbine, the reacting mixture expands and accelerates through the stator passages before transferring work to the turbine rotor. Finally, the exhaust flow is discharged through the turbine--nozzle mixing plane at high velocity to generate thrust. Overall, the predicted distributions of mixture fraction, total temperature, and Mach number are consistent with the characteristic compression, combustion, expansion, and exhaust processes of a small gas-turbine engine.

\begin{figure}[H]
\centering
\includegraphics[width=0.95\textwidth]{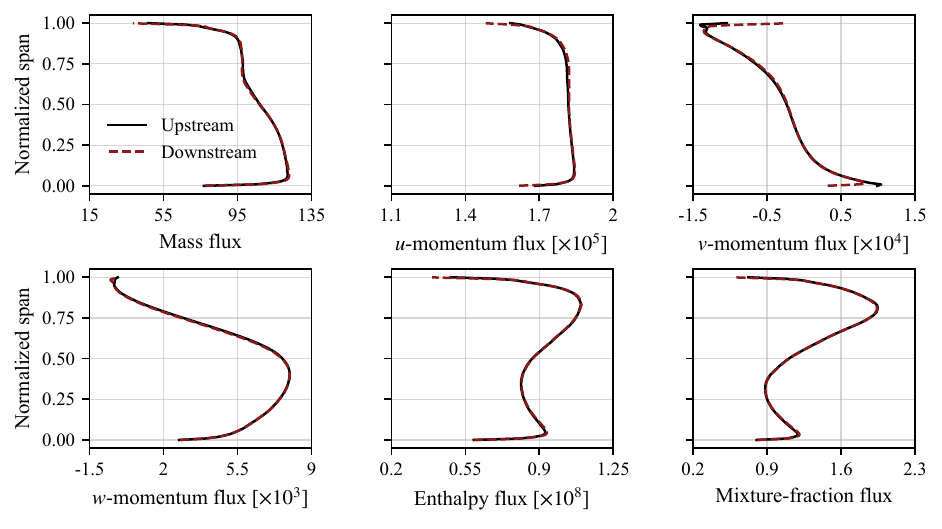}
\caption{Radial distribution of fluxes at the combustor--turbine mixing plane at $1\times10^5$ rpm.}
\label{fig:kj66_flux}
\end{figure}

Regarding the flux conservation, Figure~\ref{fig:kj66_flux} shows the radial distribution of circumferential area-averaged mass, momentum, enthalpy and flamelet variable fluxes upstream and downstream of the combustor--turbine mixing plane. Despite the thermochemical coupling, the radial distributions of the conserved fluxes on both sides of the mixing plane closely overlap. This conservation behaviour follows directly from the flux-based mixed-out reconstruction. Unlike primitive variables averaging, the present method first integrates the mass, three-component momentum, total-enthalpy, and scalar fluxes at the upstream side of the interface, and then reconstructs a single downstream mixed-out state that reproduces these conservative quantities. The upstream and downstream radial flux distributions therefore confirm that the proposed method preserves conservative transport under reacting-mixture conditions, rather than merely matching averaged primitive variables.

\begin{figure}[H]
\centering
\includegraphics[width=0.95\columnwidth]{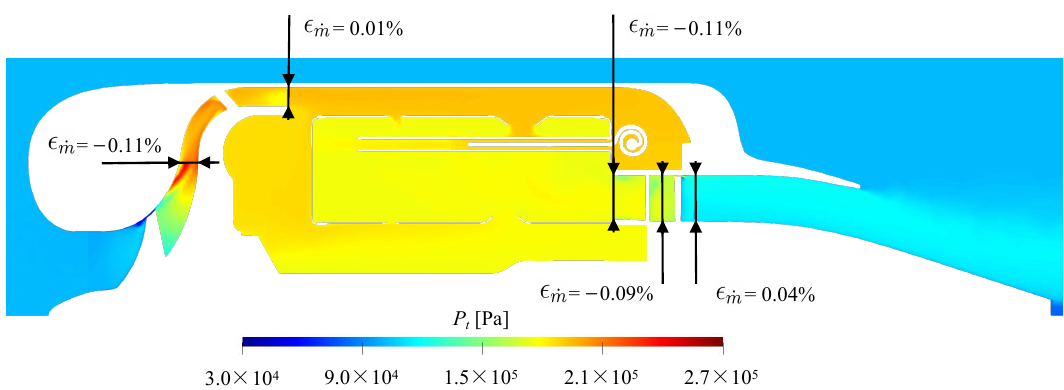}
\caption{Mass flow conservation across the five KJ66 mixing planes at $1\times10^{5}$ rpm, showing the total pressure field and relative upstream–downstream mass flow errors.}
\label{fig:kj66_pt}
\end{figure}

The mass flow rates at distinct mixing planes are shown in Figure~\ref{fig:kj66_pt}. The relative error $\epsilon_{\dot{m}}$ is calculated as the difference between the upstream and downstream mass flow rates at the mixing plane, normalised by the upstream mass flow rate. Even under the present complex flow conditions, where the total pressure varies by more than one order of magnitude across the engine, the maximum relative error at each mixing plane remains below $0.11\%$. This indicates that mass flow conservation is well preserved by the proposed reacting-mixture mixing-plane method in the whole aero-engine simulation.

\begin{figure}[ht]
\centering
\includegraphics[width=0.95\columnwidth]{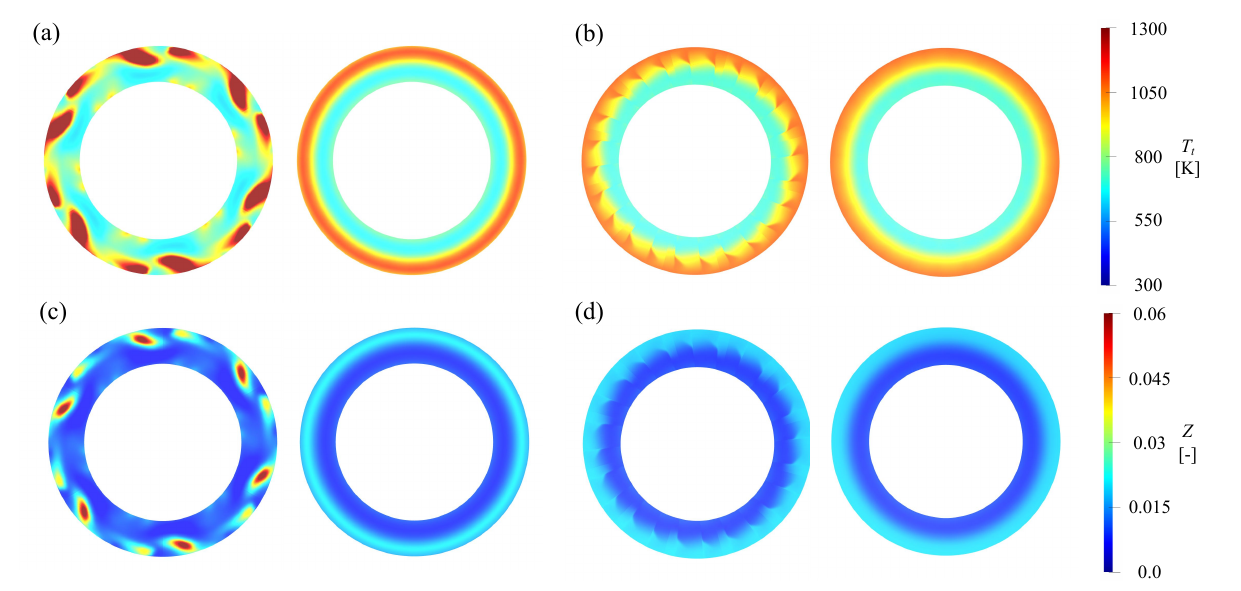}
\caption{Comparison of the upstream circumferential fields (left) and reconstructed mixed-out states (right) at two mixing planes: total temperature at (a) the combustor–turbine and (b) turbine stator–rotor interfaces, and mixture fraction at (c) the combustor–turbine and (d) turbine stator–rotor interfaces.}
\label{fig:kj66_flowfield4}
\end{figure}

The combustor exit is the location where thermochemical effects have the strongest influence on the gas properties. Therefore, the proposed mixing-plane method demonstrates its main value at the combustor--turbine interface, where the transferred flow involves strong variations in temperature, composition, and thermophysical properties. The upstream and downstream contour distributions at this interface are shown in Figure~\ref{fig:kj66_flowfield4}(a). The hot-streak structure generated at the combustor exit is circumferentially mixed and transferred to the turbine stator inlet. Although deterministic unsteady phenomena are removed by the steady mixing-plane treatment, conservation of total-enthalpy flux provides a physically reasonable radial distribution of total temperature at the turbine stator inlet. Meanwhile, conservation of scalar flux ensures that the downstream gas properties are evaluated from the reconstructed reacting-mixture state, rather than from an inappropriate fixed-composition air model. At the next downstream mixing plane, namely the turbine stator--rotor interface, the working fluid remains a high temperature reacting mixture, as shown in Figure~\ref{fig:kj66_flowfield4}(b). Since only $Z=0$ corresponds to pure air, the nonzero mixture fraction distribution at this interface confirms that the working fluid is still a post-combustion reacting mixture.

These results further demonstrate the thermochemical consistency of the proposed interface treatment. At reacting-mixture interfaces, the mixed-out state must satisfy both the conservative flux constraints and the mixture equation of state. In the present formulation, the scalar flux determines the reconstructed mixture state, after which the temperature, enthalpy, gas constant, and heat capacity are evaluated consistently from the same thermochemical closure. This avoids the inconsistency that may arise when these quantities are averaged independently. As a result, the downstream turbine row receives a mixed-out state that is both conservative and compatible with the reacting-mixture thermodynamic model, thereby maintaining conservative transfer across the components..

\begin{figure}[htbp]
\centering
\includegraphics[width=0.9\columnwidth]{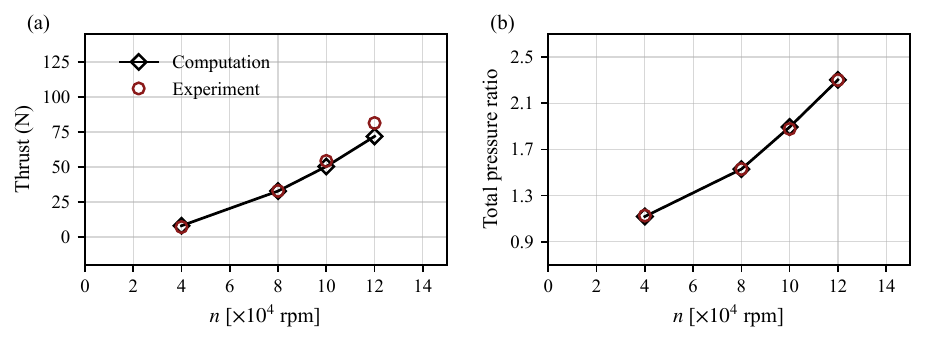}
\caption{Comparison of predicted and measured KJ66 performance as functions of rotational speed: (a) thrust and (b) total pressure ratio.}
\label{fig:kj66_overall_performance}
\end{figure}

To evaluate the effectiveness of the steady whole-engine simulation, thrust and total pressure ratio are selected as the primary performance metrics for comparison with experimental data. In addition to the design operating condition, simulations are performed at rotational speeds of 40,000, 80,000, and 120,000 rpm. The corresponding results are presented in Figure~\ref{fig:kj66_overall_performance}. The computed results show good agreement with the experimental measurements over the investigated speed range, indicating that the proposed mixing-plane method enables reliable prediction of whole-engine performance.

\FloatBarrier

\section{Conclusions}
\label{sec:conclusions}

In this paper, a generalised mixing-plane formulation has been developed for steady RANS simulation of compressible reacting flows. The method overcomes the limitation of conventional mixing-plane formulations on fixed thermodynamic properties, and allows thermodynamically consistent information transfer across interfaces containing reacting mixtures, combustion products, and thermochemical property variations. At the same time, the classical flux-conservative formulation is recovered as the perfect-gas limiting case.

The proposed method is based on a two-stage state recovery algorithm to overcome the nonlinearity caused by the energy equation. The outer iteration determines the mixed-out pressure that satisfies the EOS closure, whereas the inner iteration reconstructs the temperature through enthalpy inversion under the reacting-mixture thermochemical closure. A robust pressure-search strategy was further introduced to ensure reliable convergence when the normal velocity approaches zero.

Three test cases have been carried out to demonstrate the performance of the approach. First, a one-dimensional theoretical test was used to show that the proposed reacting-mixture algorithm preserves the imposed conservative fluxes to round-off accuracy for different mixture fraction profiles and fuel species. The fuel-dependence study further indicated that the local stiffness of the enthalpy inversion is associated with the temperature dependence of the mixture heat capacity. For flow conditions where $u_n \approx 0$, roots were successfully recovered when the pressure sampling was sufficiently concentrated near the admissible upper pressure limit.
The Darmstadt transonic compressor case verified the non-reacting limit of the formulation. When the scalar field was prescribed as $Z=0$, the reconstructed composition reduced to fixed-composition air, and the proposed method recovered the behaviour of a conventional mixing plane. This confirms that the method does not introduce artificial thermochemical effects in computations with a perfect gas.
Finally, the KJ66 micro turbine engine simulation was used to demonstrate the capability of the method in a whole-engine configuration involving both non-reacting and reacting-mixture interfaces. The proposed method evaluated the corresponding thermophysical properties from the reconstructed mixture state and maintained conservative transfer across the interfaces. The maximum relative mass flow error among the mixing planes remained below $0.11\%$, and the predicted thrust and total pressure ratio showed good agreement with available experimental data over the investigated speed range. This demonstrates the suitability of the current approach for steady multiphysics whole-engine simulations.

\bibliographystyle{elsarticle-num}
\bibliography{cas-refs}
\end{document}